\documentstyle[12pt]{article}

\def\bbga{B_{\al}\,^{\ga\de}}

\def\bib{\bibitem}
\def\ccga{C_{\al}\,^{\ga\de}}

\def\ega{e_{\al}\,^{\ga}}

\def\dem{\det e^{-1}}

\def\hga{H_{\al\be}\,^{\ga}}

\def\intf{\int d^{4}x\,}

\def\lar{\longrightarrow}

\def\pa{\partial}

\def\psiba{\overline{\psi}}
\def\tr{\,\mbox{tr}\,}
\def\Tr{\,\mbox{Tr}\,}
\def\al{\alpha}
\def\be{\beta}
\def\ga{\gamma}
\def\de{\delta}
\def\ep{\varepsilon}
\def\ze{\zeta}

\def\la{\lambda}

\def\Si{\Sigma}
\def\va{\varphi}
\def\om{\omega}

\def\beq{\begin{equation}}
\def\eeq{\end{equation}}
\def\bed{\begin{displaymath}}
\def\eed{\end{displaymath}}
\def\beqq{\begin{eqnarray}}
\def\eeqq{\end{eqnarray}}
\def\bedd{\begin{eqnarray*}}
\def\eedd{\end{eqnarray*}}

\begin{document}

\centerline{\normalsize\bf A POINCARE GAUGE THEORY OF GRAVITATION}
\baselineskip=16pt
\centerline{\normalsize\bf IN MINKOWSKI SPACETIME}

\vspace*{0.9cm}
\centerline{\footnotesize C. WIESENDANGER}
\baselineskip=13pt
\centerline{\footnotesize\it School of Theoretical Physics,
Dublin Institute for Advanced Studies}
\baselineskip=12pt
\centerline{\footnotesize\it 10 Burlington Road, Dublin 4, Ireland}
\centerline{\footnotesize E-mail: wie@stp.dias.ie}

\vspace*{0.9cm}
\baselineskip=13pt
\abstract{The conventional role of spacetime geometry in the description
of gravity is pointed out. Global Poincar$\acute{\mbox{e}}$ symmetry as an inner
symmetry of field
theories defined on a fixed Minkowski spacetime is discussed. Its extension
to local {\bf P\/} gauge symmetry and the corresponding {\bf P\/} gauge fields
are
introduced. Their minimal coupling to matter is obtained. The scaling 
behaviour of
the partition function of a spinor in {\bf P\/}
gauge field backgrounds is computed. The corresponding renormalization 
constraint is
used to determine a minimal gauge field dynamics.}

\normalsize\baselineskip=15pt

\section{Introduction}

The great success of modern particle physics in the description of the 
microscopic
interactions of elementary particles relies much on the concept of gauge field
theories. Only within this framework it was possible to formulate consistent
renormalizable quantum field theories for interacting fields in four dimensions
\cite{lor}.
This provided a strong motivation to review the gravitational interaction and 
its
beautiful description in the general theory of relativity both at the classical
and the quantum level from the point of view of gauge field theories \cite{iva}.

One central task thereby is to identify the correct gauge group for gravity.
To settle this question rather different propositions were made in the
literature on gauge approaches to gravity and yielded much new insight in the
structure of gravitational interactions \cite{iva}. The earliest attempt was 
centered about
the Lorentz group {\bf L\/} \cite{uti} and was enlarged to an analysis of 
the full
Poincar$\acute{\mbox{e}}$ group {\bf P\/} only a few years later \cite{kib}. 
Since
then there were many other contributions based on {\bf P\/} \cite{hay2}-
\cite{kik2},
the translation group {\bf T\/} \cite{hay1}-\cite{kik1} or even larger groups,
e.g. \cite{yan}-\cite{lord}.

Another important task is to analyze the status of the spacetime manifold in the
description of gravity \cite{gri}, \cite{pet}. On one hand gauge theories in 
elementary particle physics
may be quantized without spoiling renormalizability only on Minkowski 
spacetime. 
As soon as a non-trivial background geometry is introduced the usual 
perturbative
approaches face serious difficulties \cite{buc}. On the other hand the 
description of gravity
in the general theory of relativity is just given in geometrical terms 
affecting the
structure of spacetime itself. Accordingly one may ask whether some new 
light is shed
on the problems in quantizing gravity using a complementary description which
disentangles the structure of spacetime from gravitational physics.

To obtain such a description it is important to notice the purely 
conventional role
of spacetime geometry in the description of the behaviour of matter pointed 
out by
Poincar$\acute{\mbox{e}}$ already \cite{poi}. In fact two equivalent points 
of view
are possible \cite{mit}.

Either, one defines the line element $ds^{2}$ to be of Minkowskian form. 
Accordingly,
in a gravitational field material rods will shrink and clocks slow down 
w.r.t. this
metric. Hence, one defines the geometry of spacetime to be Minkowskian 
whereas the
behaviour of physical rods and clocks has to be determined by experiments.

Or, one defines rods or clocks to have one and the same length or period at any
point of spacetime. Accordingly, a measurement of the line element $ds^{2}$ 
using
these rods and clocks will yield that spacetime is curved in general. This 
is the
convention Einstein introduced to describe gravitation.

The general theory of relativity and its extensions are based on the second 
point of
view \cite{sexu} which is very natural as long as one is interested in the
macroscopic aspects of gravitation \cite{str}. Its limitation shows up at 
the quantum level. To
extend a picture so intimately related to classical concepts such as rods and 
clocks
to a simple microscopic understanding of gravitation is very difficult. In
microphysics spacetime geometry enters only as a background concept necessary in
defining a field theory. It cannot be subject to direct measurements in this 
context.

Hence, at the quantum level one is naturally led to the first point of view 
avoiding
the problematic interrelation of spacetime structure and gravitational 
phenomena. Here
free matter is described by local, causal fields defined on Minkowski 
spacetime and
its interactions are introduced using the gauge principle which allows a 
far-reaching
generalization of the connection between conservation laws and global symmetry
requirements \cite{lor}.

To obtain a gauge theory of gravitation we ensure the conservation of
energy-momentum and angular momentum by the requirement of global covariance 
of the
free matter field theory under the Poincar$\acute{\mbox{e}}$ group. We give a
complementary formulation of {\bf P\/} symmetry and its consequences
in the form of an inner symmetry (section 2) suggesting an analogy to the
description of the action of inner symmetry groups as groups of generalized
'rotations' in field space \cite{lor}. In particular the coordinate system 
used to
specify the spacetime events is not affected anymore by {\bf P\/} 
transformations.

We next introduce local {\bf P\/} gauge transformations and demand the 
invariance
of physical processes under those (section 3). This necessarily leads to the
existence of gauge fields with definite behaviour under local {\bf P\/} gauge
transformations. Their coupling to any other field is essentially fixed as 
in the
case of other gauge field theories (section 4).

To restrict the classical gauge field dynamics we demand consistency with
renormalization properties of matter fields. In a renormalizable theory the
anomalous contribution to a change of a matter partition function under 
rescaling
may be absorbed in the classical actions for the different fields (e.g. 
\cite{wip}).
As an example we determine the change of the one-loop partition function for 
a Dirac
spinor under rescaling (sections 5 and 6) and may accordingly fix a minimal 
gauge field
action (section 6).

We work on Minkowski spacetime ({\bf R\/}$^{4}$,$\eta$) with Cartesian 
coordinates
throughout, such that $\eta=\mbox{diag}(1,-1,-1,-1)$. Indices $\al,\be,\ga,...$
from the first half of the Greek alphabet denote quantities defined on
({\bf R\/}$^{4}$,$\eta$) which transform covariantly w.r.t. the Lorentz group.
They are correspondingly raised and lowered with $\eta$.

\section{Global Poincar$\acute{\mbox{e}}$ symmetry as an inner symmetry}

Let us consider a set of fields $\va_{j}(x)$ with $j=1,..,n$ which are 
defined on
Minkowski spacetime ({\bf R\/}$^{4}$,$\eta$) and belong to some representations
of the Poincar$\acute{\mbox{e}}$ group. Their dynamics shall be specified by the
action 
\beq \label{1} S_{M}=\intf\,{\cal L}_{M}(\va_{j},\pa_{\al}\va_{j}) \eeq
where the Lagrangian ${\cal L}_{M}$ is assumed to be real.

The usual conception of global Poincar$\acute{\mbox{e}}$ transformations, 
acting partly
on spacetime and partly in inner field space, is expressed in the 
transformation formulae
\beqq \label{6} x^{\al}\lar x'^{\al}&=&x^{\al}+\ep^{\al}+\om^{\al}\,_{\be}
x^{\be},
\nonumber \\ \va_{j}(x)\lar \va_{j}'(x')&=&\va_{j}(x)-\frac{i}{4}\om^{\al\be}
\Si_{\al\be}\,\,\va_{j}(x). \eeqq
$\de x^{\al}=\ep^{\al}+\om^{\al}\,_{\be}x^{\be}$ is the change of $x$ under the
combination of a global infinitesimal spacetime translation and an infinitesimal
Lorentz rotation,
$\de\va_{j}=-\frac{i}{4}\om^{\al\be}\Si_{\al\be}\,\,\va_{j}(x)$ the 
corresponding change of $\va_{j}$
in field space. $\Si_{\ga\de}$ are the representations of the generators of 
the Lie
algebra {\bf so(1,3)\/} in inner field space normalized to fulfil the 
commutation
relations
\beq [\Si_{\ga\de},\Si_{\ep\zeta}]=2i\{\eta_{\de\ep}\Si_{\ga\zeta}-
\eta_{\de\zeta}
\Si_{\ga\ep}+\eta_{\ga\ep}\Si_{\zeta\de}-\eta_{\ga\zeta}\Si_{\ep\de}\}. \eeq

If the action Eq. (\ref{1}) does not change under the continuous transformations
Eqs. (\ref{6}) the field theory is globally Poincar$\acute{\mbox{e}}$ invariant.
Accordingly, Noether's theorem yields a conserved vector quantity
\beq \label{8} J^{\ga}=\Theta^{\ga}\,_{\al}\cdot\ep^{\al}+\frac{1}{2}
{\cal M}^{\ga}
\,_{\al\be}\cdot\om^{\al\be}. \eeq
As $\ep^{\al}$ and $\om^{\al\be}$ vary independently the canonical energy 
momentum tensor
\beq \Theta^{\ga}\,_{\al}=\frac{\pa{\cal L}_{M}}{\pa(\pa_{\ga}\va_{j})}\cdot
\pa_{\al}\va_{j}-\eta^{\ga}\,_{\al}\cdot{\cal L}_{M} \eeq
and the canonical angular momentum tensor
\beq {\cal M}^{\ga}\,_{\al\be}=\Theta^{\ga}\,_{\al}x_{\be}-\Theta^{\ga}\,_{\be}
x_{\al}+\frac{i}{2}\frac{\pa{\cal L}_{M}}{\pa(\pa_{\ga}\va_{j})}
\Si_{\al\be}\,\va_{j} \eeq
are individually conserved.

As a complementary conception we now introduce global infinitesimal {\bf P\/} 
gauge
transformations
\beqq \label{10} x^{\al}\lar x'^{\al}&=&x^{\al} \nonumber \\
\va_{j}(x)\lar \va_{j}'(x)&=&\Bigl(({\bf 1}+\Theta)\va_{j}\Bigr)(x) \eeqq
where the infinitesimal hermitean gauge operators are given by
\beq \Theta=-\{\ep^{\ga}+\om^{\ga\de}x_{\de}\}\pa_{\ga}
-\frac{i}{4}\om^{\ga\de}\Si_{\ga\de}. \eeq
Much in analogy to non-abelian gauge field theory {\bf P\/} acts as a Lie group
of generalized 'phase rotations' $({\bf 1}+\Theta)$ in field space only and 
leaves the
spacetime coordinates $x$ unchanged. Note that one can also decompose $\Theta$
w.r.t. the the {\bf p\/} algebra generators $p_{\ga}=i\,\pa_{\ga}$ and
$m_{\ga\de}=i(x_{\ga}\pa_{\de}-x_{\de}\pa_{\ga})+\frac{1}{2}\Si_{\ga\de}$
which emphasizes the aforementioned useful analogy even more \cite{wie2}.

Eqs. (\ref{10}) define again a symmetry transformation of globally
Poincar$\acute{\mbox{e}}$ invariant actions as the corresponding Lagrangian just
picks up a total divergence under a {\bf P\/} gauge transformation
\beqq & &{\cal L}_{M}(\va_{j}'(x),\pa_{\al}\va_{j}'(x))=
{\cal L}_{M}(\va_{j}(x),\pa_{\al}\va_{j}(x))  \nonumber \\
& &\quad-\{\ep^{\ga}+\om^{\ga}\,_{\be}x^{\be}\}\cdot\pa_{\ga}
{\cal L}_{M}(\va_{j}(x),\pa_{\al}\va_{j}(x)) \eeqq
which does not contribute to the action integral.

Hence, we are led to a complementary conception of Poincar$\acute{\mbox{e}}$ 
symmetry
as a purely inner symmetry. The corresponding Noether symmetry current is 
found to be
the same $J^{\ga}$ as in Eq. (\ref{8}). This shows that the two global 
conceptions are
equivalent w.r.t. their physical consequences. On the other hand it is 
conceptually
easy to generalize global {\bf P\/} gauge transformations to local ones and 
to build up
the corresponding gauge theory in analogy to the non-abelian case because 
the structure
of spacetime and the action of the gauge group on the fields remain strictly 
separated.

\section{Local P gauge invariance and the covariant derivative 
${\tilde\nabla}_{\al}$}

Let us extend {\bf P\/} to a Lie group of local infinitesimal gauge 
transformations
by allowing $\ep(x)$ and $\om(x)$ to vary with $x$. We thus consider from now on
\beq \label{18} \Theta(x)=-\{\ep^{\ga}(x)+\om^{\ga\de}(x)x_{\de}\}\pa_{\ga}
-\frac{i}{4}\om^{\ga\de}(x)\Si_{\ga\de}. \eeq
Note that the algebra of the $\Theta(x)$'s does close again. There is a new 
element of
non-commutativity in their algebra as, contrary to the usual case, the
local parameters $\ep(x)$ and $\om(x)$ do not commute with the generators 
$\pa_{\ga}$
of the algebra. The emerging ordering problem is overcome by the
convention that $\Theta(x)$ in its above form only acts to the right. This 
convention
is motivated by demanding equivalence of the algebra of the $\Theta(x)$'s to the
diffeomorphism times {\bf so(1,3)\/} algebra. The formulae (\ref{10}) still 
define the
representation of {\bf P\/} in the space of fields.

Next, to recast a given matter theory in a locally {\bf P\/} gauge invariant 
form
we must introduce a covariant derivative ${\tilde\nabla}_{\al}$ which is 
defined by the
requirement
\beq \label{20} {\tilde\nabla}'_{\al}\,\left({\bf 1}+\Theta(x)\right)
=\left({\bf 1}
+\Theta(x)\right)\,{\tilde\nabla}_{\al}. \eeq
Here ${\tilde\nabla}'_{\al}$ denotes the gauge transformed covariant derivative.
Because ${\tilde\nabla}_{\al}$ transforms as a Lorentz vector we have to
supplement the generators $\Si_{\ga\de}$ 
of {\bf so(1,3)\/} in matter field space 
occurring in the decomposition of $\Theta(x)$ with the corresponding generators
$\Si_{\ga\de}$ acting 
on vectors to obtain the appropriate product representation as
we will automatically do wherever necessary from now on.

We find that the Lagrangian with covariant derivatives ${\tilde\nabla}_{\al}$
replacing the usual 
ones behaves under local infinitesimal {\bf P\/} transformations as
\beqq \label{19} & &{\cal L}_{M}(\va_{j}'(x),{\tilde\nabla}'_{\al}\va_{j}'(x))=
{\cal L}_{M}(\va_{j}(x),{\tilde\nabla}_{\al}\va_{j}(x)) \nonumber \\
& &\quad-\{\ep^{\ga}(x)+\om^{\ga\de}(x)x_{\de}\}\cdot\pa_{\ga}
{\cal L}_{M}(\va_{j}(x),{\tilde\nabla}_{\al}\va_{j}(x)). \eeqq
Note that this does 
not yet ensure the local {\bf P\/} invariance of the original
action $S_{M}=\int{\cal L}_{M}$ 
because the second term in Eq. (\ref{19}) is no longer a
pure divergence as it was in the case of global infinitesimal transformations.

Now, to fulfil Eq. (\ref{20}) we use the ansatz 
\beq \label{21} \pa_{\al}\lar {\tilde\nabla}_{\al}=
\ega\pa_{\ga}+\frac{i}{4}\bbga\Si_{\ga\de} \eeq
decomposing 
${\tilde\nabla}_{\al}$ w.r.t. $\pa_{\ga}$ and $\Si_{\ga\de}$ in the same
way as the local gauge operator $\Theta(x)$ in Eq. (\ref{18}). This ansatz is
compatible with 
the required behaviour Eq. (\ref{20}) of the covariant derivative
under local 
{\bf P\/} gauge transformations provided the 16 compensating translation
gauge fields $\ega$ transform as
\beqq \label{30} \de\ega&\equiv&e'_{\al}\,^{\ga}-\ega
=e_{\al}\,^{\ze}\cdot\pa_{\ze}\{\ep^{\ga}+\om^{\ga\de}x_{\de}\} \nonumber \\
&-&\{\ep^{\ze}+\om^{\ze\eta}x_{\eta}\}\cdot\pa_{\ze}\ega
+\om_{\al}\,^{\ze}e_{\ze}\,^{\ga} \eeqq
and the 24 Lorentz gauge fields $\bbga$ as
\beqq \label{31} \de\bbga&\equiv&B'_{\al}\,^{\ga\de}-\bbga
=e_{\al}\,^{\ze}\cdot\pa_{\ze}\om^{\ga\de}
-\{\ep^{\ze}+\om^{\ze\eta}x_{\eta}\}\cdot\pa_{\ze}\bbga \nonumber \\
&+&\om_{\al}\,^{\ze}B_{\ze}\,^{\ga\de}+\om^{\ga}\,_{\ze}B_{\al}\,^{\ze\de}
+\om^{\de}\,_{\ze}B_{\al}\,^{\ga\ze}. \eeqq

As in our conception coordinate and {\bf P\/} gauge transformations are strictly
separated we emphasize that the introduction of $\ega$ and $\bbga$ has neither
implications on the structure of the underlying spacetime which we assumed to be
({\bf R\/}$^{4}$,$\eta$) endowed with the Minkowski metric $\eta$. Nor has it
implications 
on the maximal symmetry group of ({\bf R\/}$^{4}$,$\eta$), which is the
Poincar$\acute{\mbox{e}}$ 
group if we still restrict ourselves to the use of Cartesian
coordinates only.

With the abbreviations
\beq d_{\al}\equiv\ega\pa_{\ga},\quad\quad
B_{\al}\equiv\frac{i}{4}\bbga\Si_{\ga\de},\eeq
where $\Si_{\ga\de}$ 
must be properly adjusted to the Lorentz group representation
it acts upon, we write 
\beq {\tilde\nabla}_{\al}=d_{\al}+B_{\al}\eeq
from now on. $d_{\al}$ is just the translation covariant derivative introduced
in \cite{wie}. We finally remark that ${\tilde\nabla}_{\al}$ may alternatively
be decomposed w.r.t. the {\bf p\/} algebra generators $p_{\ga}$ and
$m_{\ga\de}$ yielding 
the most convenient starting point for perturbation expansions
\cite{wie2}.

\section{The field strength operator. Minimal coupling to matter fields}

Before turning 
to the field strength operator itself we introduce the non {\bf P\/}
covariant translation field strength
\beq [d_{\al},d_{\be}]\equiv\hga d_{\ga} \eeq
as in \cite{wie}. $\hga$ is expressed in terms of $\ega$ as \cite{wie}
\beq \hga=e^{-1\,\ga}\,_{\ep}(e_{\al}\,^{\ze}\cdot\pa_{\ze} e_{\be}\,^{\ep}-
e_{\be}\,^{\ze}\cdot\pa_{\ze} e_{\al}\,^{\ep}) \eeq
where $e^{-1\,\ga}\,_{\ep}$ is the matrix inverse to $e_{\al}\,^{\ep}$, i.e.
$e_{\al}\,^{\ep}\cdot e^{-1\,\ga}\,_{\ep}=\de_{\al}\,^{\ga}$.

This allows us 
now to obtain the field strength operator and its decomposition in a
simple way. Taking 
into account the vector character of ${\tilde\nabla}_{\al}$
a little algebra yields
\beqq S_{\al\be}\equiv [{\tilde\nabla}_{\al},{\tilde\nabla}_{\be}]
&=&\hga d_{\ga}-(B_{\al\be}\,^{\ga}-B_{\be\al}\,^{\ga})d_{\ga} \nonumber \\
&+&d_{\al}B_{\be}-d_{\be}B_{\al}+[B_{\al},B_{\be}]. \eeqq
Introducing the tensor coefficients of $d_{\ga}$
\beq \label{36} T_{\al\be}\,^{\ga}\equiv B_{\al\be}\,^{\ga}-B_{\be\al}\,^{\ga}
-H_{\al\be}\,^{\ga} \eeq
we may rewrite $S_{\al\be}$ as
\beq \label{37} [{\tilde\nabla}_{\al},{\tilde\nabla}_{\be}]=
-T_{\al\be}\,^{\ga}\,
{\tilde\nabla}_{\ga}+\frac{i}{4}{\tilde R}^{\ga\de}\,_{\al\be}\Si_{\ga\de}, \eeq
where ${\tilde R}^{\ga\de}\,_{\al\be}$ is found to be
\beqq \label{38} {\tilde R}^{\ga\de}\,_{\al\be}&\equiv&
d_{\al}B_{\be}\,^{\ga\de}-d_{\be}B_{\al}\,^{\ga\de}+B_{\al}\,^{\de\ep}
B_{\be\ep}\,^{\ga} \nonumber \\ &-&B_{\be}\,^{\de\ep} B_{\al\ep}\,^{\ga}
-H_{\al\be}\,^{\ep} B_{\ep}\,^{\ga\de}. \eeqq
As $S_{\al\be}$ has a 
decomposition w.r.t. ${\tilde\nabla}_{\de}$ and $\Si_{\ga\de}$
it acts in general not only as a matrix but also as a first order differential
operator in field space.

From Eq. (\ref{36}) we see that only if $\bbga$ is related
to $H_{\al\be}\,^{\ga}$ 
the tensor $T_{\al\be}\,^{\ga}$ may vanish. Denoting this
particular $\bbga$ with $\ccga$ the required relation becomes
\beq C_{\al\be}\,^{\ga}-C_{\be\al}\,^{\ga}=H_{\al\be}\,^{\ga}. \eeq
We may now solve for $\ccga$ in terms of $H_{\al\be}\,^{\ga}$ with the result
\beq \ccga=\frac{1}{2}\left(H_{\al}\,^{\ga\de}-H_{\al}\,^{\de\ga}
-H^{\ga\de}\,_{\al}\right). \eeq
Whenever $\bbga=\ccga$, 
i.e. $T_{\al\be}\,^{\ga}=0$ we omit the tilde, hence writing
\beq \label{42} \nabla_{\al}\equiv d_{\al}+C_{\al}. \eeq
Obviously we obtain now for $S_{\al\be}$ a matrix only
\beq \label{43} [\nabla_{\al},\nabla_{\be}]=\frac{i}{4}R^{\ga\de}\,_{\al\be}
\Si_{\ga\de}. \eeq

By construction $S_{\al\be}$ transforms homogeneously under infinitesimal local
{\bf P\/} gauge transformations
\beq S'_{\al\be}\left({\bf 1}+\Theta(x)\right)=\left({\bf 1}+\Theta(x)\right)
\,S_{\al\be} \eeq
leading to
\beq \de T_{\al\be}\,^{\ga}=\Theta(x)\,T_{\al\be}\,^{\ga},\quad\quad
\de {\tilde R}^{\ga\de}\,_{\al\be}=
\Theta(x)\,{\tilde R}^{\ga\de}\,_{\al\be}. \eeq
$T_{\al\be}\,^{\ga}$ and 
${\tilde R}^{\ga\de}\,_{\al\be}$ transform homogeneously
under infinitesimal local {\bf P\/} gauge transformations. We emphasize that
$T_{\al\be}\,^{\ga}=0$ is indeed a gauge covariant statement as we implicitly
assumed above introducing $\ccga$.

It is very convenient 
to introduce the homogeneously transforming difference of the
two gauge fields
\beq \label{49} K_{\al}\,^{\ga\de}\equiv \bbga-\ccga \eeq
which is related to $T_{\al\be}\,^{\ga}$ as
\beq K_{\al\be}\,^{\ga}-K_{\be\al}\,^{\ga}=T_{\al\be}\,^{\ga} \eeq
with the obvious inversion
\beq K_{\al}\,^{\ga\de}=\frac{1}{2}\left(T_{\al}\,^{\ga\de}-T_{\al}\,^{\de\ga}
-T^{\ga\de}\,_{\al}\right). \eeq
We can express now ${\tilde R}^{\ga\de}\,_{\al\be}$ in terms of
$R^{\ga\de}\,_{\al\be}$, which we take as our fundamental
field strength variable, and $K_{\al}\,^{\ga\de}$ as
\beqq  {\tilde R}^{\ga\de}\,_{\al\be}&=&R^{\ga\de}\,_{\al\be}
+\nabla_{\al}K_{\be}\,^{\ga\de}-\nabla_{\be}K_{\al}\,^{\ga\de} \nonumber \\
&+&K_{\al}\,^{\de\ep} K_{\be\ep}\,^{\ga}
-K_{\be}\,^{\de\ep} K_{\al\ep}\,^{\ga}. \eeqq

Next, let us 
turn to the extension of globally {\bf P\/} invariant matter actions
to locally gauge invariant ones. In the previous section we have obtained the
covariant derivative 
${\tilde\nabla}_{\al}$ yielding the behaviour Eq. (\ref{19})
of a matter field 
Lagrangian under local {\bf P\/} gauge transformations which is not
yet sufficient for the original action $S_{M}=\int {\cal L}_{M}$ to be locally
{\bf P\/} gauge invariant. Completing the Lagrangian with $\dem$
\beq \label{52} \dem\cdot{\cal L}_{M}(\va_{j},{\tilde\nabla}_{\al}\va_{j}) \eeq
we find 
that the combination Eq. (\ref{52}) changes under a local {\bf P\/} gauge
transformation by a pure divergence only
\beqq & &\det e'^{-1}\cdot{\cal L}_{M}(\va_{j}',{\tilde\nabla}'_{\al}\va_{j}')=
\dem\cdot{\cal L}_{M}(\va_{j},
{\tilde\nabla}_{\al}\va_{j}) \nonumber \\ & &\quad
-\pa_{\ga}\left(\{\ep^{\ga}(x)+\om^{\ga\de}(x)x_{\de}\}\dem\cdot
{\cal L}_{M}(\va_{j},{\tilde\nabla}_{\al}\va_{j})\right). \eeqq

Therefore the minimally extended locally {\bf P\/} gauge invariant matter action
finally becomes
\beq \label{56} S_{M}=\intf \dem(x)\cdot{\cal L}_{M}(\va_{j}(x),
{\tilde\nabla}_{\al}\va_{j}(x)). \eeq
Of course, $S_{M}$ 
remains invariant if we change from one to another inertial system
by global coordinate translations or Lorentz rotations.

It is the conception of 
{\bf P\/} symmetry as an inner symmetry together with the
gauge principle which has led us to this minimal coupling prescription. In this
conception the gauge fields and their transformation behaviour do not interfere
with the spacetime 
structure ({\bf R\/}$^{4}$,$\eta$) fixed by an a priori convention
and the underlying geometry remains separated from the physics described by the
{\bf P\/} gauge fields in the same manner it remains separated from the physics
described by any usual matrix gauge field.

We remark that a geometric re-interpretation of the gauge fields and their
corresponding field strengths introduced above may be given in the framework of
Riemannian geometry 
\cite{wie2}. But then the gauge group {\bf P\/} and the requirement of local
{\bf P\/} gauge invariance are replaced by the groups of general (infinitesimal)
coordinate transformations 
and local $SO(1,3)$ frame rotations and the requirement of
invariance under 
these groups \cite{sexu}, \cite{str} and the geometry of spacetime
is necessarily linked with these complementary symmetry requirements.

\section{Dirac partition function in gauge field backgrounds}

In Yang-Mills 
gauge field theory one may fix the gauge field dynamics quite uniquely
by demanding 
gauge invariance of the action and using dimensional arguments related to
the renormalization 
properties of the theory. In a similar fashion we attempt here to
obtain information 
on the classical {\bf P\/} gauge field dynamics studying quantized
matter fields and their 
renormalization properties in gauge field backgrounds. In the
Yang-Mills case our 
arguments lead straightforward to the usual Yang-Mills action.

The assumption that 
the interactions of the {\bf P\/} gauge fields with the different
matter fields are 
renormalizable imposes strong conditions on the classical gauge
field dynamics. For 
let us suppose that a given theory for a matter field and the
gauge fields $\ega$ 
and $\bbga$ is perturbatively renormalizable. Then we know that
the change of the 
partition function of the whole system under rescaling can be
absorbed in its 
classical action yielding at most a nontrivial scale dependence of
the different couplings, 
masses and wavefunction normalizations \cite{wip}. Hence, the explicit
computation of the 
change of the one-loop matter partition functions under rescaling
will allow us to constrain the classical gauge field dynamics \cite{wie}.

For brevity we restrict ourselves to the non-trivial case of a Dirac spinor
field, the analogous discussions of a scalar and a vector field theory are found
in \cite{wie2}. The globally {\bf P\/} invariant action for a Dirac spinor with
real Lagrangian is given by
\beq S_{M}=\intf\left\{ \frac{i}{2}\psiba\ga^{\al}(\pa_{\al}\psi)-\frac{i}{2}
\overline{(\pa_{\al}\psi)}\ga^{\al}\psi-m\psiba\psi\right\}. \eeq
The Dirac matrices fulfil the usual Clifford algebra 
$\{\ga_{\al},\ga_{\be}\}=2\eta_{\al\be}$ 
and the {\bf so(1,3)\/} generators become
$\Si_{\al\be}=\frac{i}{2}[\ga_{\al},\ga_{\be}]$.
The minimal extension 
prescription yields the locally {\bf P\/} gauge invariant action
\beq S_{M}=\intf\dem\left\{ \frac{i}{2}\psiba\ga^{\al}({\tilde\nabla}_{\al}\psi)
-\frac{i}{2}\overline{({\tilde\nabla}_{\al}\psi)}
\ga^{\al}\psi-m\psiba\psi\right\}. \eeq
Due to spin $\bbga$ 
enters the action. Partially integrating ${\tilde\nabla}_{\al}$
in the second term above leads to the usual form of the Dirac action
\beq \label{59} S_{M}=\intf\dem\psiba\left\{ i\ga^{\al}({\tilde\nabla}_{\al}
-\frac{1}{2}T_{\ga\al}\,^{\ga})-m\right\}\psi. \eeq
Note the occurrence of the tensor $T$ ensuring the hermiticity of the {\bf P\/}
covariant Dirac operator w.r.t.
$\left(\chi,\psi\right)_{e}=\intf\dem\,\overline{\chi}\cdot\psi$.

The spinor partition 
function in the given gauge field background is given by the
Grassmann functional integral
\beq \label{73} {\cal Z}_{\psi}[e,B]=\int{\cal D}\psiba{\cal D}\psi\,
e^{iS_{M}(\psiba,\psi;e,B)}. \eeq
Note that we omit possible normalizations in order to obtain the most general
renormalization 
structure later. As $S_{M}$ is already of the usual quadratic form
we may perform the Grassmann integral and formally obtain
\beq \label{74}{\cal Z}_{\psi}[e,B]=e^{\frac{1}{2}\log\det M_{\psi}(e,B)}. \eeq
The hyperbolic fluctuation operator $M_{\psi}(e,B)$ is obtained as usual by
squaring the Dirac operator introduced in Eq. (\ref{59}) 
\beq M_{\psi}(e,B)
\equiv-\ga^{\al}({\tilde\nabla}_{\al}-\frac{1}{2}T_{\ga\al}\,^{\ga})
\cdot\ga^{\be}({\tilde\nabla}_{\be}-\frac{1}{2}T_{\de\be}\,^{\de})-m^{2} \eeq
and is hermitean 
w.r.t. $(\,,\,)_{e}$ due to the occurrence of $T$. To make contact
to the heat kernel techniques fully described in \cite{wie2} we have to recast
$M_{\psi}$. Using $[{\tilde\nabla}_{\al},\ga^{\be}]=0$ and
$\ga^{\al}\ga^{\be}=\eta^{\al\be}-i\Si^{\al\be}$ we obtain
\beqq M_{\psi}(e,B)&=&-({\tilde\nabla}_{\al}-\frac{1}{2}T_{\ga\al}\,^{\ga})
({\tilde\nabla}^{\al}-\frac{1}{2}T_{\de}\,^{\al\de}) \nonumber \\ &+&\frac{i}{2}
\Si^{\al\be}(-T_{\al\be}\,^{\de}{\tilde\nabla}_{\de}+\frac{i}{4}
{\tilde R}^{\ga\de}\,_{\al\be}\Si_{\ga\de}-
{\tilde\nabla}_{\al}K_{\de\be}\,^{\de})-m^{2}. \eeqq
Next we write $T_{\al}\equiv\frac{i}{4}T^{\ga\de}\,_{\al}\Si_{\ga\de}$ and
absorb the first 
order derivative term $-2T_{\al}{\tilde\nabla}^{\al}$ in the second
order one. Together with the use of the Jacobi identities for the covariant
derivative ${\tilde\nabla}_{\al}$ we then find the manifestly hermitean result
\beqq M_{\psi}(e,B)
&=&-({\tilde\nabla}_{\al}+T_{\al}-\frac{1}{2}T_{\ga\al}\,^{\ga})
({\tilde\nabla}^{\al}+T^{\al}-\frac{1}{2}T_{\de}\,^{\al\de}) \nonumber \\
&+&T_{\al}\,T^{\al}
+\frac{i}{2}\Si^{\al\be}(\frac{i}{4}{\tilde R}^{\ga\de}\,_{\al\be}\Si_{\ga\de}
+{\tilde R}^{\de}\,_{\al\de\be})-m^{2}. \eeqq
This is the form relevant for further computation.

As we are interested 
in the behaviour of ${\cal Z}_{\psi}[e,B]$ under rescaling the
most suited 
renormalization of the ultraviolet divergent determinant in Eq. (\ref{74})
is based on the 
$\zeta$-function as it is a manifestly gauge invariant technique. 

One can define the 
ultraviolet regularized functional determinant of an operator $M$
satisfying certain conditions to be \cite{dow}, \cite{haw}
\beq \label{b1} \log\det M\equiv -\lim_{u\to 0}\frac{d}{du}\,\zeta(u;\mu;M) \eeq
where the generalized $\zeta$-function belonging to $M$ is given by the Mellin
transformed of the heat kernel
\beq \label{b3} \zeta(u;\mu;M)
=\frac{i\mu^{2u}}{\it\Gamma(u)}\int \limits_{0}^{\infty}
ds\,(is)^{u-1}\Tr e^{-isM}. \eeq
The scale $\mu$ at which parameters such as couplings, masses and wavefunction
normalizations have to be 
adjusted is introduced in order to keep the determinant
dimensionless.

Hence, with the use of 
Eq. (\ref{b1}) the spinor partition function normalized at scale
$\mu$ becomes
\beq {\cal Z}_{\psi}[\mu;e,B]
=e^{-\frac{1}{2}\,\zeta'(0;\mu;M_{\psi}(e,B))}. \eeq
According to the formula
\beq \label{b4} \zeta'(0;{\tilde\mu};M)=\zeta'(0;\mu;M)
+2\log\la\cdot\zeta(0;\mu;M). \eeq
we finally obtain the change of $\cal{Z}_{\psi}$ corresponding to a change of
scale ${\tilde\mu}=\la\mu$
\beq \label{82} {\cal Z}_{\psi}[{\tilde\mu};e,B]={\cal Z}_{\psi}[\mu;e,B]\cdot
e^{-\log\la\cdot\zeta(0;\mu;M_{\psi}(e,B))}. \eeq

\section{Scaling of 
the Dirac partition function and the dynamics of the gauge fields}

In the previous section we expressed the change of the one-loop spinor partition
function under rescaling in terms of the $\zeta$-function belonging to the
corresponding fluctuation 
operator. Renormalizability of any theory including dynamical
gauge fields requires 
now at least that this anomalous contribution, which is a local
polynomial in $\ega$ 
and $\bbga$ and their derivatives, may be absorbed in the classical
action for the gauge 
fields $\ega$ and $\bbga$ \cite{wip}. Hence, to determine a minimal
gauge field 
dynamics consistent with renormalizability we finally have to obtain
explicitly the value of the $\zeta$-function at zero.

In the representation 
Eq. (\ref{b3}) for $\zeta(u;\mu;M)$ it is the singular part
of the $s$-integration 
which yields a nonvanishing value for $\zeta(0;\mu;M)$. As
this singular part comes from the small $s$-region we may use the corresponding
expansion for the trace of the heat kernel \cite{wip}
\beq \Tr e^{-isM}
\quad{\sim\!\!\!\!\!\!\!\!^{s\to 0}}\frac{i}{(4\pi is)^{\frac{d}{2}}}\,
\sum_{k=0}^{\infty}(is)^{k}\int d^{d}x\dem\tr c_{k}(x) \eeq
given in terms of the 
well-known Seeley--DeWitt coefficient functions $c_{k}$. Performing
the $s$-integration in 
(\ref{b3}) singles out the contribution for $k=\frac{d}{2}$
from the infinite sum and one obtains \cite{wip}
\beq \label{b6} \zeta(0;\mu;M)=\frac{i}{(4\pi)^{\frac{d}{2}}}\,\int d^{d}x\dem
\tr c_{\frac{d}{2}}(x). \eeq

The computation of the $c_{\frac{d}{2}}$ has been done in various ways in the
literature \cite{dewi} - 
\cite{gusy} and we restrict ourselves to give the result
relevant to our case,
where $d=4$ and $\tr_{D}$ denotes the Dirac trace \cite{wie2}
\beqq \label{107} \tr_{D} c_{2}
&=&4U_{m}+(\frac{1}{6}R^{\al\be}\,_{\al\be}-m^{2})
\cdot(4V_{3}-V_{1}^{\ga\de}\,_{\ga\de}) \nonumber \\
&-&\frac{1}{6}\nabla_{\al}\nabla^{\al}(4V_{3}-V_{1}^{\ga\de}\,_{\ga\de})
-\frac{1}{24}F_{\al\be\ga\de}\cdot F^{\al\be\ga\de} \nonumber \\
&-&V_{3}\cdot V_{1}^{\ga\de}\,_{\ga\de}-\frac{1}{8}V_{2\al\be}\cdot
V_{2}^{\al\be}+\frac{1}{8}V_{1}^{\al\be}\,_{\al\be}\cdot
V_{1}^{\ga\de}\,_{\ga\de} \nonumber \\
&+&2V_{3}^{2}+\frac{1}{8}V_{1\al\be\ga\de}\cdot(V_{1}^{\al\be\ga\de}
+V_{1}^{\ga\de\al\be}+V_{1}^{\ga\be\de\al}). \eeqq
The term without $T$-dependence
\beqq \label{102} U_{m}
&=&-\frac{1}{30}\nabla_{\ga}\nabla^{\ga}R^{\al\be}\,_{\al\be}
+\frac{1}{72}R_{\al\be}\,^{\al\be}\cdot R_{\ga\de}\,^{\ga\de} \nonumber \\
&+&\frac{1}{180}R_{\al\be\ga\de}\cdot R^{\al\be\ga\de}
-\frac{1}{180}R_{\al\ga}\,^{\al}\,_{\de}\cdot R_{\be}\,^{\ga\be\de} \nonumber \\
&-&\frac{1}{6}m^{2}\cdot R^{\al\be}\,_{\al\be}+\frac{1}{2}m^{4} \eeqq
already occurs in the case of a scalar field, whereas the terms containing $T$
\beqq
V_{1\ga\de\al\be}&=&{\tilde R}_{\ga\de\al\be}
+\frac{1}{2}T_{\ga\de\eta}T_{\al\be}\,^{\eta}, \nonumber \\
V_{2\al\be}&=&\frac{1}{2}(V_{1}^{\eta}\,_{\al\eta\be}
-V_{1}^{\eta}\,_{\be\eta\al}), \nonumber \\
V_{3}&=&\frac{1}{2}{\tilde\nabla}_{\al}T_{\ga}\,^{\al\ga}
-\frac{1}{4}T_{\ga\al}\,^{\ga} T_{\de}\,^{\al\de} \eeqq
and
\beqq
F_{\ga\de\al\be}&=&{\tilde R}_{\ga\de\al\be}+{\tilde\nabla}_{\al}T_{\ga\de\be}
-{\tilde\nabla}_{\be}T_{\ga\de\al} \nonumber \\
&+&T_{\al\be}\,^{\eta}T_{\ga\de\eta}+T^{\eta}\,_{\ga\al}T_{\eta\de\be}
-T^{\eta}\,_{\ga\be}T_{\eta\de\al} \eeqq
are due to the 
spin of the Dirac field. As the result Eq. (\ref{107}) expressed in
terms of the natural variables $R,T$ becomes algebraically tedious we give it in
these variables only for the case $T=0$ \cite{chri}
\beqq \label{108} \tr_{D} c_{2}&=&\frac{1}{30}\nabla_{\ga}\nabla^{\ga}
R^{\al\be}\,_{\al\be}
+\frac{1}{72}R_{\al\be}\,^{\al\be}\cdot R_{\ga\de}\,^{\ga\de} \nonumber \\
&-&\frac{7}{360}R_{\al\be\ga\de}\cdot R^{\al\be\ga\de}
-\frac{1}{45}R_{\al\ga}\,^{\al}\,_{\de}\cdot R_{\be}\,^{\ga\be\de} \nonumber \\
&+&\frac{1}{3}m^{2}\cdot R^{\al\be}\,_{\al\be}+2m^{4}. \eeqq
Insertion of the results Eq. (\ref{107}) or Eq. (\ref{108}) into Eq. (\ref{b6})
with $d=4$ finally yields $\zeta(0;\mu;M_{\psi}(e,B))$.

We have obtained now the anomalous contribution to the rescaled spinor partition
function as a 
local {\bf P\/} gauge invariant polynomial in the fields $\ega$ and
$\bbga$, which also must 
be present in any classical gauge field dynamics consistent
with renormalizability of the spinor partition function. The analysis of the
corresponding scalar and vector field cases leads to results of the same general
structure \cite{wie2}. Hence, we 
are finally led to construct a minimal action for the gauge
fields just in terms of these {\bf P\/} gauge invariant expressions.

For $T\ne 0$ we restrict 
ourselves to the contributions of $O(\pa^{0},\pa^{2})$ 
in the derivatives and obtain as minimal classical action to this order
\beqq \label{116} S_{G}(e,B)&=&\int\dem\{{\it\Lambda}-\frac{1}{\kappa^{2}}\cdot
R_{\al\be}\,^{\al\be}+
\be_{1}\cdot T_{\ga\al}\,^{\ga} T_{\de}\,^{\al\de} \nonumber \\
&+&\be_{2}\cdot T_{\al\be\ga}T^{\al\be\ga}+
\be_{3}\cdot T_{\al\be\ga} T^{\ga\al\be}
+O(\pa^{4})\}, \eeqq
skipping possible total divergence terms. Here we have to introduce different
couplings $\kappa,\be_{1},\be_{2},\be_{3}$ 
and the constant ${\it\Lambda}$ which are
independently renormalized 
by the one-loop contribution we determined above. Note
that our reasoning 
automatically enforces a cosmological constant as to be expected
from general renormalization 
considerations. The action Eq. (\ref{116}) describes the
classical gauge field dynamics 
correctly at sufficiently low momentum scales and small
values of the couplings. 
Nevertheless, only a dynamics containing the huge number of
different $O(\pa^{4})$ 
terms as well, coming along with the same number of independent
couplings, will be 
consistent with renormalizability (see also \cite{buc}). Note that no terms
of $O(\pa^{6})$ or higher are demanded by our reasoning.

If we set $T=0$ the minimal classical action must contain the terms
\beqq \label{117} S_{G}(e)&=&\int\dem\{{\it\Lambda}-\frac{1}{\kappa^{2}}\cdot
R_{\al\be}\,^{\al\be}+
\al_{1}\cdot R_{\al\be}\,^{\al\be}\cdot R_{\ga\de}\,^{\ga\de}
\nonumber \\
&+&\al_{2}\cdot R_{\al\ga}\,^{\al}\,_{\de}\cdot R_{\be}\,^{\ga\be\de}
+\al_{3}\cdot R_{\al\be\ga\de}\cdot R^{\al\be\ga\de}\}, \eeqq
if discarding total 
divergences. The couplings $\kappa,\al_{1},\al_{2},\al_{3}$
and the constant 
${\it\Lambda}$ obtain again contributions from the one-loop scale
anomaly which has been determined above. We emphasize that $S_{G}$ is an action
for gauge fields defined on Minkowski spacetime ({\bf R\/}$^{4}$,$\eta$) and
is invariant on one hand 
under local {\bf P\/} gauge transformations, on the other
hand under global 
Poincar$\acute{\mbox{e}}$ transformations reflecting the symmetries
of the underlying spacetime.

Important aspects of 
the geometric version of the quantized theory (\ref{117}) such
as one-loop divergences and $\beta$-functions and its unitarity problems are
discussed in \cite{buc} and references given there.

\section{Conclusions}

To disentangle the 
structure of spacetime from the description of gravity we have 
given a complementary conception of Poincar$\acute{\mbox{e}}$ symmetry as a
purely inner symmetry. 
Its extension to local {\bf P\/} gauge symmetry has led us
to introduce gauge fields 
defined on a fixed Minkowski spacetime. Their coupling to
any other field has come out to be essentially fixed. We then constrained their
dynamics imposing consistency 
with renormalization properties of a Dirac field in
gauge field backgrounds. In an appropriate low energy limit the resulting gauge
field action has been shown to reduce to a form yielding the same observational
predictions as made in general relativity which confirms us to have obtained a
sensible description of gravity within the present framework.

In our conception there 
is no direct interrelation between gravity and the structure
of spacetime. Although it may 
be convenient to introduce a second, 'effective' metric
on Minkowski spacetime to 
answer questions about the behaviour of rods and clocks in
a classical context 
\cite{wie} we essentially deal here with a field theoretical
description of gravitation 
free of any non-trivial geometrical aspects as proposed 
to investigate
in the introduction. 
Unfortunately, the resulting theory is in many aspects still too
close to the geometric 
approach and is far from leading to a convincing quantization.
This shows up most clearly 
in the necessity of including terms quadratic in the field
strength in 
the classical gauge field action. Although the corresponding quantum
theory is known to be 
renormalizable, the occurrence of negative energy or negative
norm ghost states 
has destroyed up to now any attempt of establishing unitarity and
hence a physical interpretation of the theory (see also \cite{buc}).

\section*{Acknowledgments}

This work has partially been supported by Schweizerischer Nationalfonds. It is a
pleasure to thank the 
organizers of the Erice school 1995 on Quantum Gravity for the
warm and fruitful hospitality during the course. The discussions with F. W. Hehl
in Erice and his friendly interest in this work were an important encouragement.
I am indebted to C. Ford 
and F. Krahe for intensive discussions, to L. O'Raifeartaigh
for clarifying some group theoretical aspects.

\end{document}